\documentclass[preprint,12pt,3p]{elsarticle}
\usepackage{graphicx}
\usepackage{subcaption}
\usepackage{lipsum}
\usepackage{xcolor}

\title{In-situ localization of damage in a Zn-Al-Mg coating deposited on steel by continuous hot-dip galvanizing}
\author[add1,add2,cor]{Houssem Eddine Chaieb}
\author[add2]{Vincent Maurel}
\author[add2]{Kais Ammar}
\author[add2]{Samuel Forest}
\author[add3]{Alexandre Tanguy}
\author[add4]{Eva Héripré}
\author[add1]{Franck Nozahic}
\author[add5]{Jean-Michel Mataigne}
\author[add1]{Joost De Strycker}

\cortext[cor]{Corresponding author. E-mail address: houssemeddine.chaieb@arcelormittal.com (H.E. Chaieb)}
\address[add1]{ArcelorMittal Global R\&D Ghent, J.F. Kennedylaan 3, B-9060 Zelzate, Belgium}
\address[add2]{Mines Paris, PSL University, MAT - Centre des Matériaux, CNRS UMR 7633, BP 87 91003 Evry, France}
\address[add3]{CNRS UMR 7649, Laboratoire de Mécanique Des Solides, Ecole Polytechnique, Palaiseau, France}
\address[add4]{Université Paris-Saclay, CentraleSupélec, ENS Paris-Saclay, CNRS, Laboratoire de Mécanique Paris-Saclay, 91190, Gif-sur-Yvette, France}
\address[add5]{ArcelorMittal Global R\&D Maizières, Voie Romaine, Maizières lès Metz, France}

\begin{document}

\begin{abstract}
Zn-Al-Mg coatings are characterized by a complex microstructure with dendritic and eutectic phases. This heterogeneous phase distribution contributes to multiple deformation and damage mechanisms. The presence of brittle phases promotes crack initiation and propagation. This study reveals a new deformation and damage mechanism of a Zn-Al-Mg coating, \textcolor{black}{where twinning can induce crack initiation in the eutectic region}. The chronology of different events leading to crack initiation and propagation is clearly established by in-situ tensile testing in a scanning electron microscope, which helps to establish a detailed characterization of the mechanical behavior of the coating.
\end{abstract}


\maketitle





The increasing use of zinc-based coatings is justified by their high corrosion resistance as well as their excellent mechanical properties such as wear resistance and adhesion \cite{evy,parisotpart1}. Zinc-aluminum-magnesium coatings represent a significant part of the production of galvanized coatings, especially in the automotive industry and in the production of solar panels. They are characterized by the presence of different phases, including brittle phases \cite{AHMADI2020108364,AHMADI2021103041,Nozahic}. The latter are preferential sites for cracking and can significantly affect the structural integrity of the coating. 

The mechanical behavior of zinc-based coatings has been studied in several previous works \cite{parisotpart1,AHMADI2020108364, AHMADI2022142995, LEGENDRE2019138156,Song2,Park2017}. The respective influence of composition, microstructure and crystallographic texture were related to the formability of the coatings. However, the deformation and damage mechanisms of Zn-Al-Mg coatings have not been thoroughly investigated. Such coatings are characterized by the presence of dendritic and eutectic phases, which makes them significantly different from pure zinc coatings or coatings with low alloying element content ($ < $ 0.2 wt\%). Previous work has focused on the role of the binary eutectic phase present in Zn-Al-Mg coatings and its contribution to the formability of the coating: it was found to be a brittle phase in which most cracks preferentially form \cite{AHMADI2020108364,AHMADI2021103041,AHMADI2022114453,AHMADI2021110215}.

The present work provides new insights into the chronology of deformation and damage events in Zn-Al-Mg coatings. This allows for a better understanding of how cracks are initiated and what can be done to further optimize the microstructure and improve the formability of the coatings. In particular, a new damage mechanism resulting from the interaction between dendrites and eutectic phases is revealed. Indeed, a causal interaction between these two phases is observed, which affects the deformation and damage mechanisms of the coating. Special attention is paid to the contribution of each phase to the crack propagation in 3D in order to clarify the crack growth mechanisms through the thickness of the coating.

\begin{figure*}[h!]
	\includegraphics[scale=0.45]{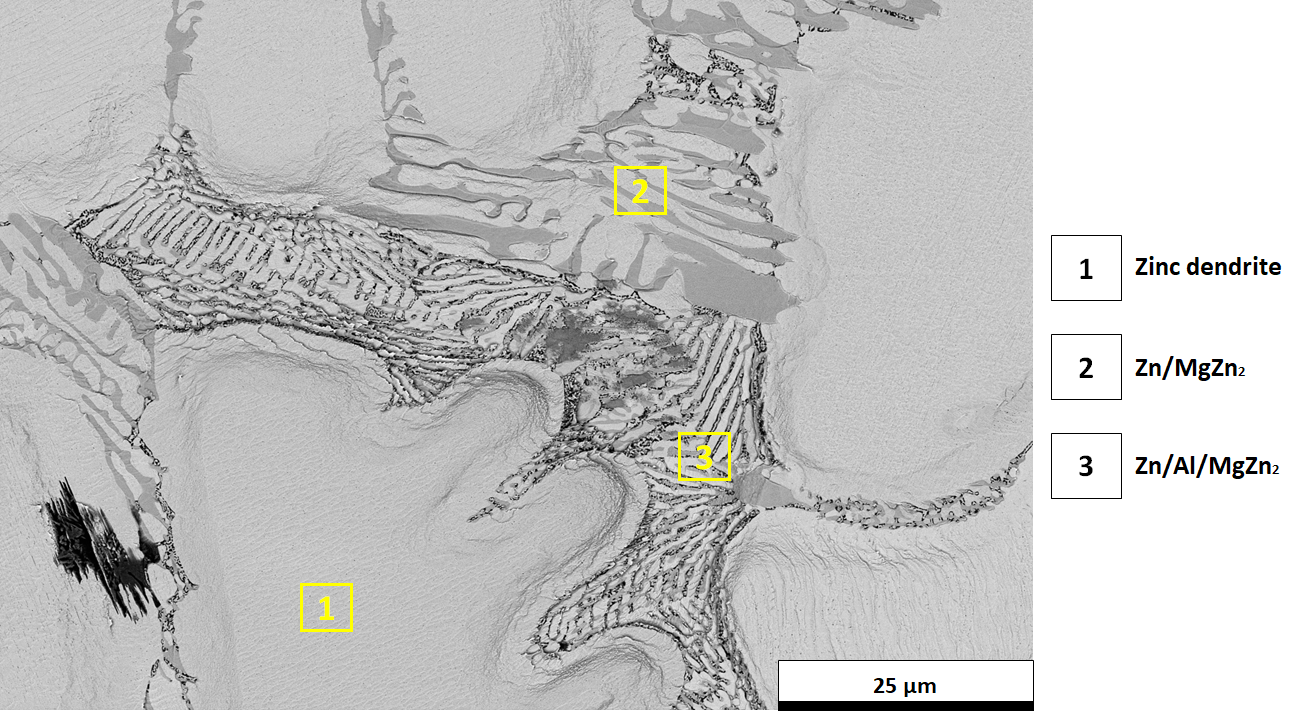}
	\centering
	\caption{SEM image showing the microstructure of the Zn-1.2Al-1.2Mg coating in backscattered electrons: Area 1 shows a zinc dendrite, area 2 is the binary eutectic phase Zn/MgZn$_2$ and area 3 is the ternary eutectic phase Zn/Al/MgZn$_2$, the sample is prepared by ion polishing.}
	\label{mic}
\end{figure*}

In this work, an industrial Zn-Al-Mg coating deposited on steel by continuous hot-dip galvanizing is studied. \textcolor{black}{The steel composition and mechanical properties are shown in Table \ref{SteelCompo} and Table \ref{MechProp} respectively.} The bath composition of the coating is Zn-1.2\%Al-1.2\%Mg. The microstructure of the coating is shown in Figure \ref{mic}. The coating is composed of pro-eutectic zinc dendrites and two eutectic phases: a coarse binary eutectic Zn/MgZn$_2$ and a thin ternary eutectic Zn/Al/MgZn$_2$. The dendritic phase is predominant with an area fraction of about 67\% as determined by surface image analysis. The film thickness is 5 µm.

\begin{table}[h!]
    \centering
    \caption{Chemical composition (wt\%) of the steel substrate}
    \begin{tabular}{ccc ccc}
    \hline
        C & Mn & P & S & Si & Ti\\ \hline
        $\le0.12$ & $\le0.6$ & $\le0.1$ & $\le0.045$ & $\le0.5$ & $\le0.3$\\ \hline
    \end{tabular}
    \label{SteelCompo}
\end{table}

\begin{table}[h!]
    \centering
    \caption{Mechanical properties of the steel substrate}
    \begin{tabular}{cc}
    \hline
        Rm - Tensile Strength (MPa) & 270 - 330\\ \hline
        ReH - Yield Strength (MPa) & 170 - 180\\ \hline
        A - Elongation L0=80 mm (\%) & 41\\ \hline
    \end{tabular}
    \label{MechProp}
\end{table}

\begin{figure}[h!]
  \begin{subfigure}[b]{0.6\columnwidth}
    \includegraphics[width=0.85\linewidth]{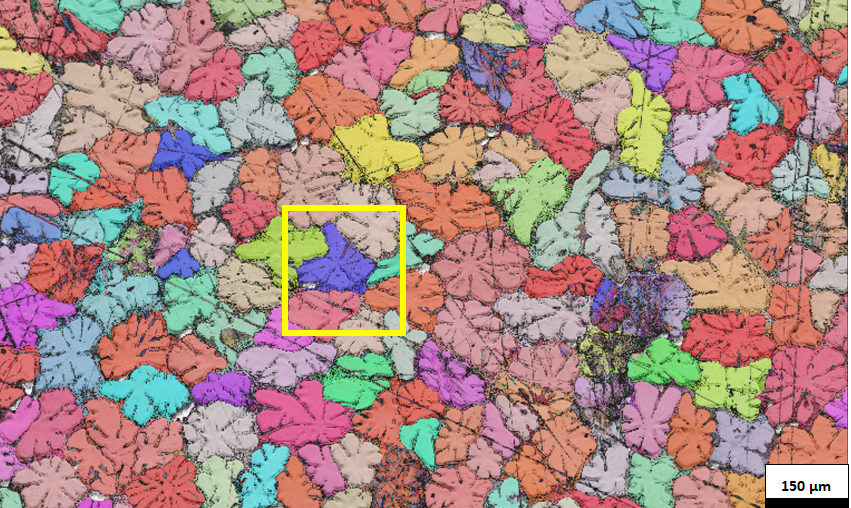}
    \caption{IPF-Z map of the Zn-1.2Al-1.2Mg coating}
    \label{fig:1}
  \end{subfigure}
  \hfill 
  \begin{subfigure}[b]{0.35\columnwidth}
    \includegraphics[width=\linewidth]{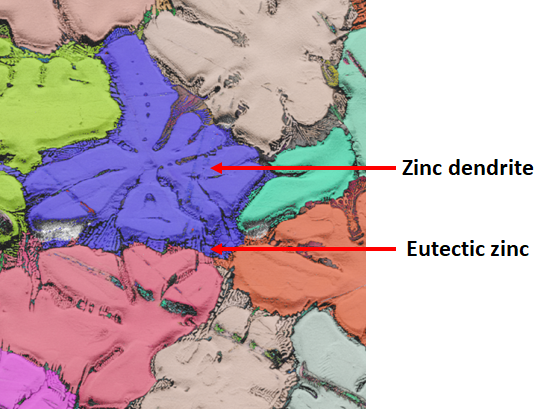}
    \caption{A detail of the area highlighted in Figure \ref{fig:1} showing the epitaxy relationship between proeutectic zinc dendrites and eutectic zinc}
    \label{fig:2}
  \end{subfigure}
  
    \begin{subfigure}[b]{0.7\columnwidth}
    \includegraphics[width=\linewidth]{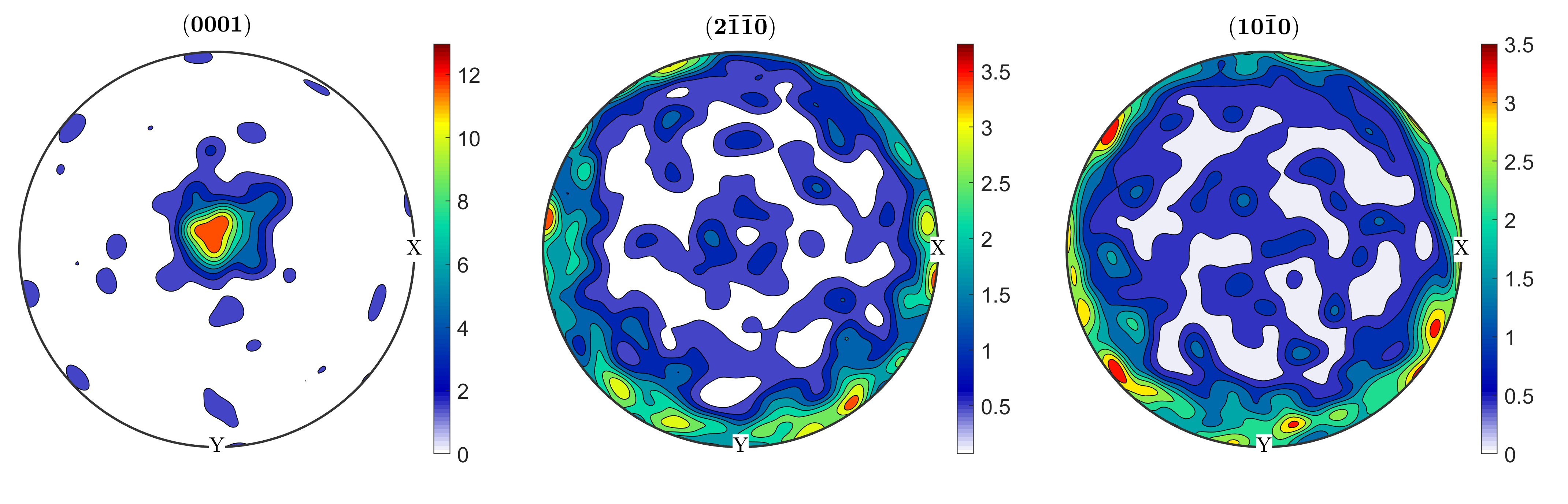}
    \caption{Pole figures of area analyzed in Figure \ref{fig:1}}
    \label{fig:3}
  \end{subfigure}
  \hfill 
  \begin{subfigure}[b]{0.25\columnwidth}
    \includegraphics[width=\linewidth]{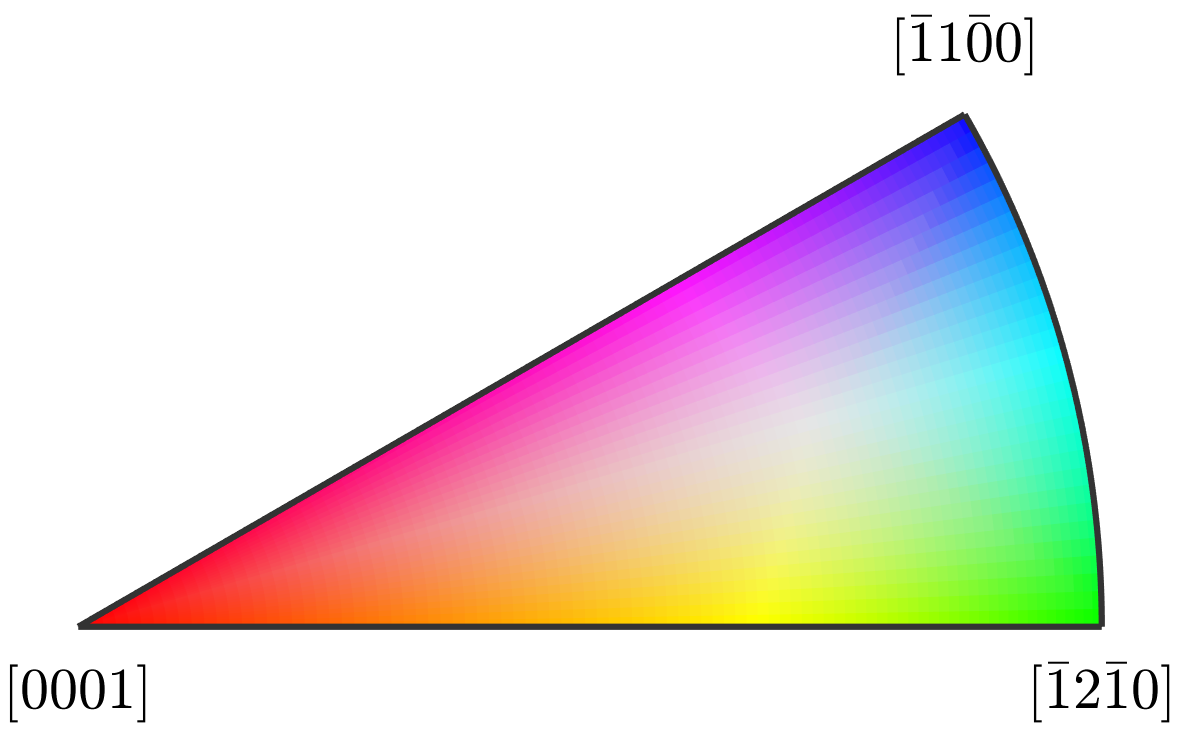}
    \caption{The standard IPF triangle}
    \label{fig:4}
  \end{subfigure}
  
      \caption{EBSD analysis of Zn-1.2Al-1.2Mg coating, Z is normal to the coating surface}
    \label{Ebsd}
\end{figure}

The crystallographic texture of the coating is studied using electron backscatter diffraction (EBSD) (Figure \ref{Ebsd}). It is recalled that zinc has a hexagonal close packed (HCP) structure. The coating is characterized by a strong basal orientation (Figure \ref{fig:3}), meaning that for most grains the c-axis of the hexagonal cell is perpendicular to the surface of the coating. However, some grains have different orientations, which can change the mechanical behavior of the coating (Figure \ref{fig:2}) \cite{AHMADI2022114453}. An important finding is the epitaxial relationship between the proeutectic zinc dendrites and the eutectic zinc located in their vicinity (Figure \ref{fig:2}). In other words, the eutectic zinc has the same orientation as the adjacent primary zinc dendrite, which can also affect the behavior of the coating.

\begin{figure}[h!]
  \begin{subfigure}[b]{0.3\columnwidth}
    \includegraphics[width=\linewidth]{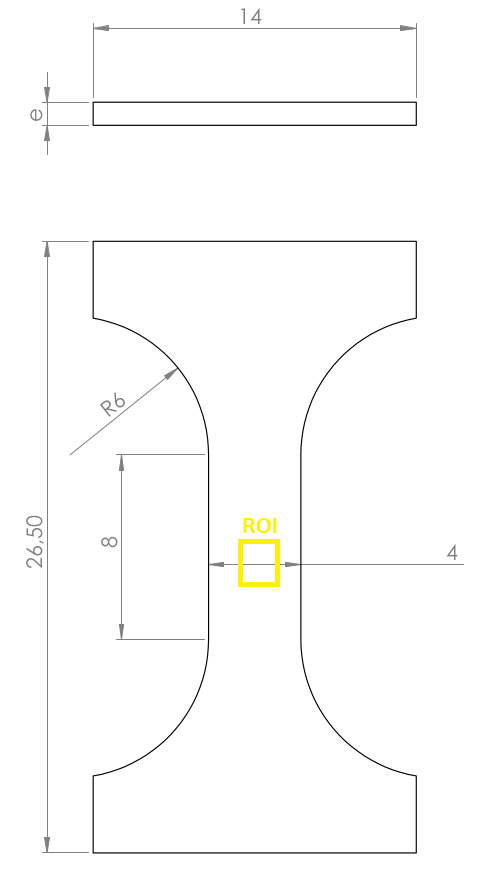}
    \caption{SEM in-situ tensile specimen (dimensions in mm)}
    \label{specimen}
  \end{subfigure}
  \hfill 
  \begin{subfigure}[b]{0.6\columnwidth}
    \includegraphics[width=\linewidth]{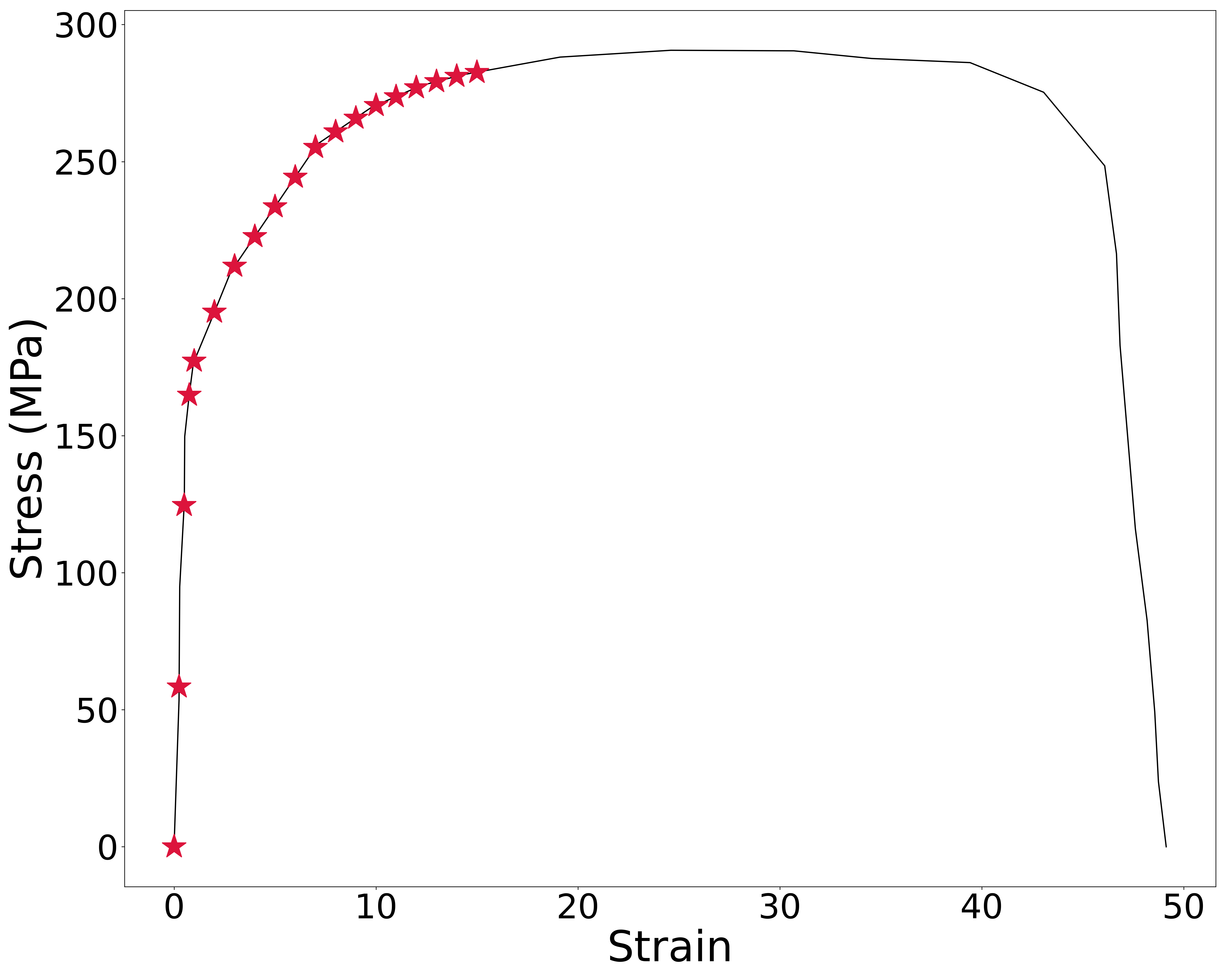}
    \caption{Stress-strain curve of the coated steel substrate. The markers highlight the strain levels for which the SEM images are taken during the in-situ tensile test}
    \label{curve}
  \end{subfigure}
  \begin{center}
        \begin{subfigure}[b]{1\columnwidth}
    \includegraphics[width=1\linewidth]{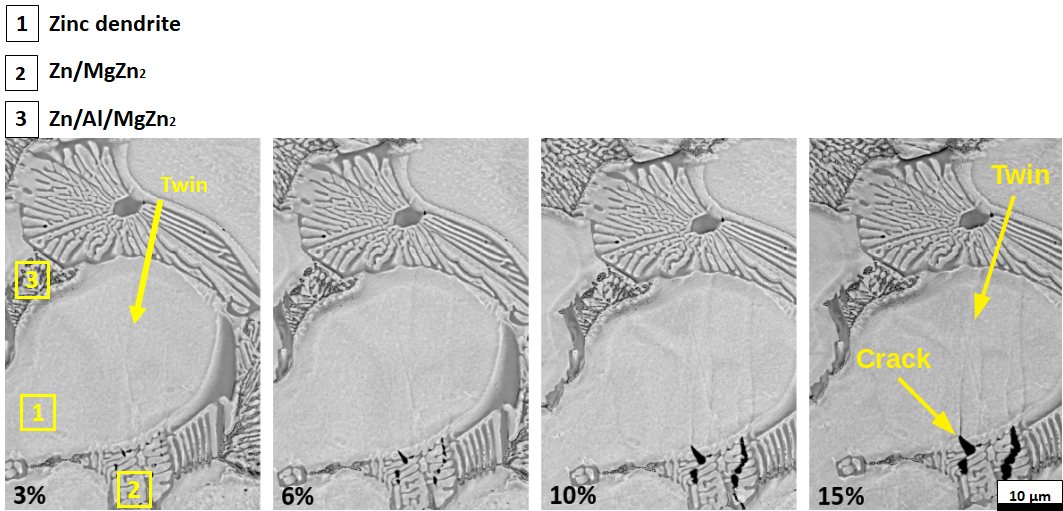}
    \caption{Chronology of crack formation in the coating, the macroscopic strain level is shown at the corner of each step, tensile loading direction is horizontal}
    \label{dam}
  \end{subfigure}
  \end{center}

      \caption{SEM in-situ study of deformation and damage mechanisms in Zn-1.2Al-1.2Mg coating}
    \label{damage}
\end{figure}

\begin{figure}[h!]
\begin{center}
    
  \begin{subfigure}[b]{0.5\columnwidth}
    \includegraphics[width=0.9\linewidth]{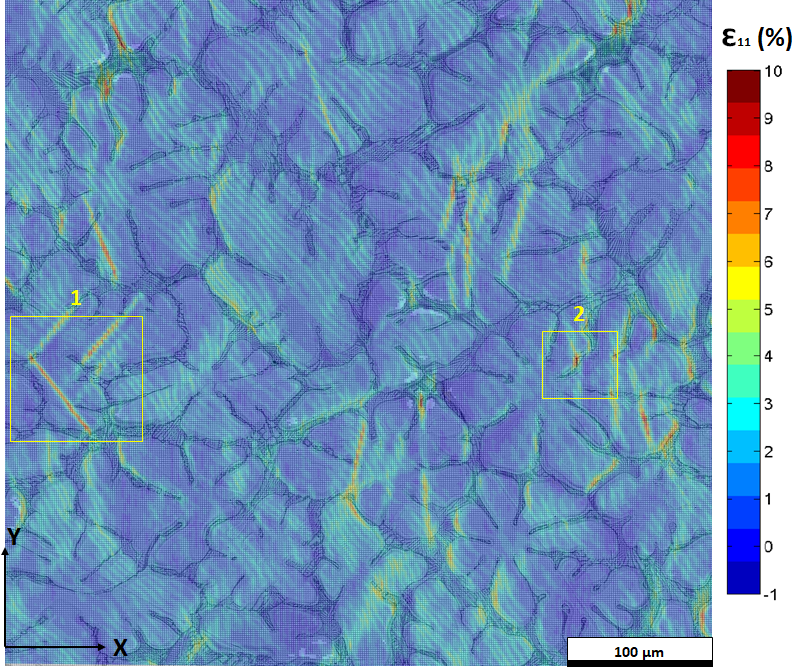}
    \caption{Strain field measurements}
    \label{dic1}
  \end{subfigure}
  \hfill 
  \begin{subfigure}[b]{0.4\columnwidth}
    \includegraphics[width=\linewidth]{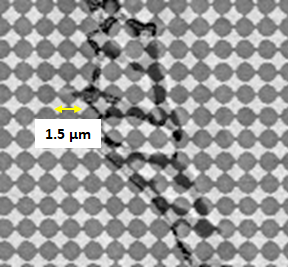}
    \caption{A detail of the gold pattern used for DIC}
    \label{dic2}
  \end{subfigure}
  
    \begin{subfigure}[b]{0.5\columnwidth}
    \includegraphics[width=0.8\linewidth]{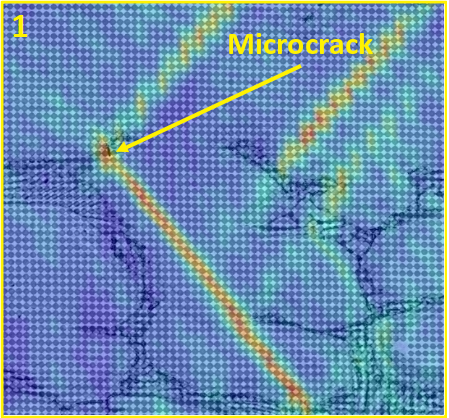}
    \caption{Detail of area 1 in Figure \ref{dic1}}
    \label{dic3}
  \end{subfigure}
  \hfill 
  \begin{subfigure}[b]{0.4\columnwidth}
    \includegraphics[width=\linewidth]{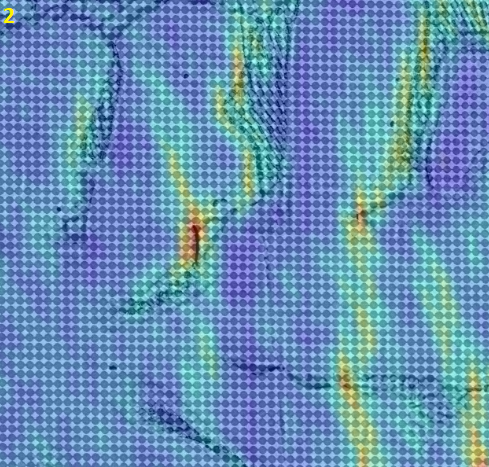}
    \caption{Detail of area 2 in Figure \ref{dic1}}
    \label{dic4}
  \end{subfigure}
  \end{center}
      \caption{Strain field measurements using DIC (1\% of macroscopic strain) showing the relationship between the twinning mechanism and crack formation within the binary eutectic phase, tensile direction is horizontal.}
    \label{dic}
\end{figure}

An in situ dog bone tensile specimen was cut from an industrial product processed at ArcelorMittal.  The surface was prepared using a Gatan PECS II ion polisher. An in-situ SEM tensile test was then performed up to 15\% macroscopic strain. Images were taken for each 1\% increment of global strain. \textcolor{black}{The stress-strain curve corresponding to the coated substrate as well as the strain values where the SEM images were taken are shown in Figure \ref{curve}}. The area of the specimen gauge is 8 $\times$ 4 mm$^2$ with a thickness of 0.8 mm for the substrate (Figure \ref{specimen}). Within this gauge, a region of interest (ROI) corresponding to 1.5 $\times$ 1.5 mm$^2$ was defined on which gold patterns were deposited by electron beam lithography \cite{ALLAIS19943865} for further digital image correlation (DIC) analysis (Figure \ref{dic2}).

Figure \ref{dam} details an area where a crack is observed within a dendrite. This crack is located at twin boundaries. In fact, a twin band is observed at 3\% macroscopic strain. The latter grows with increasing strain and induces microcracking of the binary eutectic phase (inside MgZn$_2$). Starting from 6\% macroscopic strain, this microcrack extends in the eutectic zinc of the binary eutectic phase and then grows towards the zinc dendrite along the twin boundaries. At higher strain levels this crack propagates and opens.
\textcolor{black}{The formation of cracks within the binary eutectic phase can occur without necessarily being associated the formation of a twin in an adjacent primary dendrite, but the occurrence of such a phenomenon promotes the cracking within this phase. This finding is well in line with the results reported in previous works \cite{AHMADI2020108364,AHMADI2021103041, AHMADI2022142995} where the  MgZn$_2$ phase was identified as the most brittle phase and the most prone to cracking and where the role of twinning has been identified as detrimental for similar coatings with coarse grain structure.
The crystallographic orientation of the grains as well as the steel substrate condition the cracking behaviour of the coating \cite{AHMADI2022114453,AHMADI2021110215}, but this topic is beyond the scope of this short paper. However, the results showing the role of twinning in the formation of cracks within the binary eutectic phase and at the twin boundaries suggest that a crystallographic orientation with less basal grains could improve the cracking behaviour of the coating by drastically reducing the number of grains susceptible to twinning and eventually the number of transgranular cracks.}

To better understand this mechanism, strain fields were calculated using the DIC technique (Figure \ref{dic}). This is done using the CMV software \cite{Bornert}. Twinning is observed within a zinc dendrite at early stages of strain. This twinning induces a high local level of strain that causes the fracture of the brittle MgZn$_2$ phase, which belongs to the binary eutectic phase (Figure \ref{dic3}). Such a mechanism is often observed in this study and is thought to be the main cause of cracking in the binary eutectic. However, it should be noted that some other microcracks are observed within the binary eutectic phase without being caused by a twin.

The mechanism of mutual cracking between pro-eutectic zinc and the binary eutectic phase is unique and is observed for the first time on coatings. The initiation of cracks at twin boundaries for steel has been discussed in \cite{twin}, but no previous work has dealt with this mechanism for coatings or established a link between the phases. The specificity of the mechanism studied in this work is that it shows the influence of each phase on the other phases: the twin initially formed is responsible for crack initiation within the binary eutectic phase. The same crack grows and induces crack initiation at the twin boundaries inside the zinc dendrites. We also recall the fact that the basal orientation is dominant in the coating, meaning that most of the grains have an orientation that promotes twinning, making this mechanism very likely to occur. Furthermore, the epitaxial relationship between the pro-eutectic zinc dendrites and the eutectic zinc justifies the extension of twinning bands within the binary eutectic phase.

\textcolor{black}{Concerning strain localization, it is noteworthy that interfaces between phases play a key role, Figure \ref{dic}. This could be explained by the heterogeneity of the micro-mechanical properties of the  phases involved. For example, the primary zinc dendrite shows a very different behaviour compared to the binary eutectic phase. This aspect is well described in a previous work \cite{AHMADI2020108364} where the binary eutectic was shown to be the most detrimental constituent in terms of micro-ductility, despite  having a higher yield strength compared to primary zinc and ternary eutectic. The presence of the brittle intermetallic compound MgZn$_2$ is the main cause for such a difference in micro-mechanical properties.}

\textcolor{black}{The strain localization at the interface could also be explained by the difference in topography between the phases. In fact, the primary zinc dendrites occupy a plateau region while the eutectic phases represent a valley. This contrast in topography and height could make the interfaces a preferential site for strain localization.}

\begin{figure}[h!]
\begin{center}
\centering
  \begin{subfigure}[b]{\columnwidth}
  \begin{center}
    \includegraphics[width=0.6\linewidth]{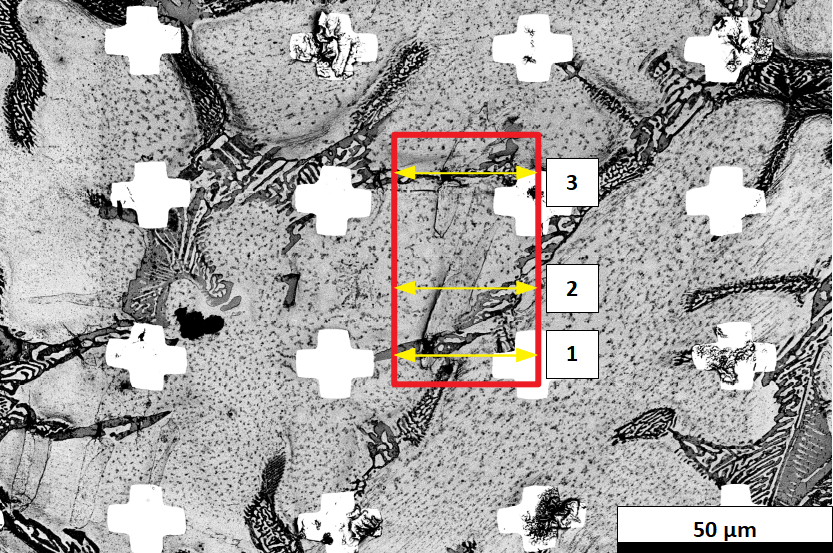}
    \caption{BSE image of the ROI where the cracking mechanism is observed}
    \label{fib1}
     \end{center}
  \end{subfigure}
   \end{center}
\begin{center}
     \begin{subfigure}[b]{0.6\columnwidth}
    \includegraphics[width=\linewidth]{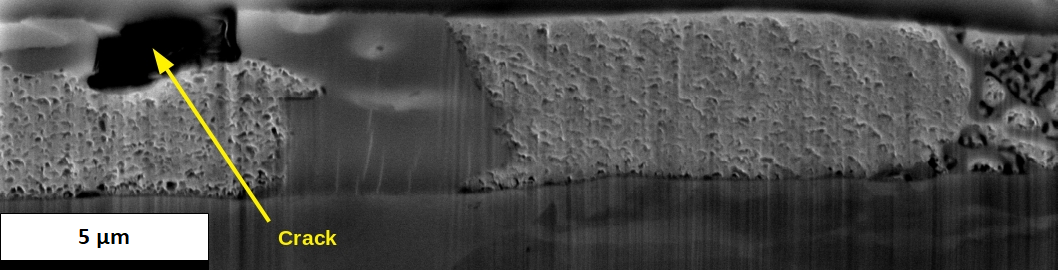}
    \caption{Cross-section, see arrow 1 in Figure \ref{fib1}}
    \label{fib3}
  \end{subfigure}
  \hfill 
  \begin{subfigure}[b]{0.6\columnwidth}
    \includegraphics[width=\linewidth]{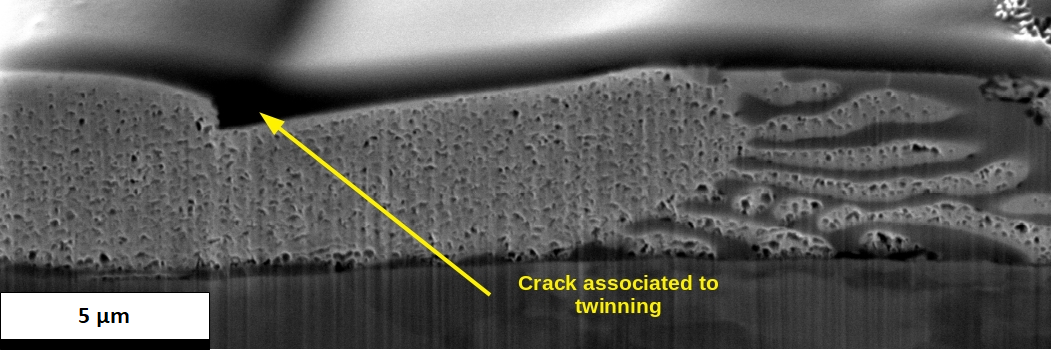}
    \caption{Cross-section, see arrow 2 in Figure \ref{fib1}}
    \label{fib4}
  \end{subfigure}
    \hfill 
    \begin{subfigure}[b]{0.6\columnwidth}
    \includegraphics[width=\linewidth]{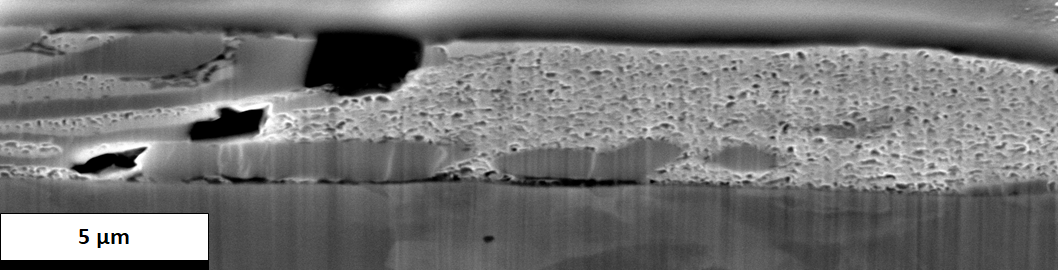}
    \caption{Cross-section, see arrow 3 in Figure \ref{fib1}}
    \label{fib5}
  \end{subfigure}
  \end{center}
      \caption{3D FIB analysis of Zn-1.2Al-1.2Mg coating}
    \label{FIB}
\end{figure}

A more detailed study of the observed crack is performed using SEM - Focused Ion Beam (SEM-FIB), where the 3D Slice\&View technique is used \cite{TRUGLAS2020110407} (Figure \ref{FIB}). \textcolor{black}{To observe the extent of the crack in the coating thickness, several sections are made in an area where this mechanism is observed}. Figure \ref{fib1} shows some of the cross-sections. We observe that cracks forming within the binary eutectic phase (MgZn$_2$ phase) can reach the steel-coating interface in a discontinuous manner (Figure \ref{fib5}). Zinc dendrites, eutectic zinc, and the ternary eutectic act as barriers to these cracks (Figure \ref{fib3}). As for the crack growth at the twin boundaries and initially induced by the fracture of the binary eutectic phase, it remains superficial (Figure \ref{fib4}): this type of cracks is associated with a slight perturbation of the local roughness of the coating without further propagation towards the steel substrate.

\textcolor{black}{This work complements the recently  published works  that have  studied the mechanical behavior of Zn-Al-Mg coatings. It is clear that twinning plays a crucial role in conditioning the deformation and the damage mechanisms of such coatings. This was highlighted in \cite{AHMADI2022142995} where the effect of the grain size of a Zn-Al-Mg coating on its formability was discussed. It was shown that a coating with a coarse microstructure would have a decrease in its mechanical performance and this is mainly due to the presence of large transgranular cracks caused by twinning. This aligns with the findings of the current work. This also gives an indication of a potential improvement path. In fact, a coating with a more refined microstructure should perform better because the presence of more grains leads to the presence of more grain boundaries that act as barriers to crack propagation. In addition, limiting the grain size should limit the occurrence of twinning  and the size of the twins,  resulting in an optimized mechanical robustness. Another work \cite{AHMADI2022114453} highlighted the role of the crystallographic texture where a strong basal orientation was found to be a factor in improving formability. This contradicts with the previous statements since a strong basal orientation  leads to a higher twinning rate. Therefore, it has become clear that there is a certain trade-off between the grain size and the crystallographic texture that needs to be defined: the grain size should be  optimized first then the actions towards having a fiber-like basal orientation should be undertaken.} 

In conclusion, we have investigated a new deformation and damage mechanism for the Zn-1.2Al-1.2Mg coating where a mutual causal effect between the pro-eutectic zinc dendrites and the binary eutectic phase is observed. Twinning causes high local strain levels leading to crack initiation within the brittle MgZn$_2$ phase. Such cracks can grow through the zinc dendrite along the twin boundaries. Some other microcracks are observed within the binary eutectic phase. The eutectic zinc and the ternary eutectic phase were observed to act as barriers to crack propagation. The MgZn$_2$ is the most detrimental phase in the coating. These observations give an idea of what could be done to improve the formability of the coating: reducing the volume fraction of the binary eutectic phase is essential for better formability. The 3D observations showed that the coating has an overall good adhesion to the steel substrate. Some cracks can reach the steel-coating interface in a discontinuous manner and form mainly within the MgZn$_2$ phase.

This work is part of a more comprehensive analysis of the mechanical behavior of hot-dip Zn-Al-Mg coatings that will be presented in future work.


\section*{Acknowledgments}

This work was carried out in collaboration with several laboratories (Centre des Matériaux, LMS X, LMPS, OCAS NV). The FIB-SEM work was carried out using the facilities available at the LMPS laboratory within the MATMECA consortium, supported by the ANR under contract number ANR-10-EQPX-37. The contribution of Fabrice Gaslain, Gwyneth Vandeborne and Dimitri Monteyne is highly appreciated. Special thanks are due to Alain Koster who supervised some of the mechanical tests related to this work.

\section*{Declaration of Competing Interest}
The authors declare that there is, to their knowledge, no financial interests that could influence the research work presented in this paper.

\section*{Funding}

This research work was performed as a part of the PhD "Multiscale approach of mechanical behavior of hot-dip Zn-Al-Mg coatings deposited on steel sheet", performed at Centre des Matériaux of MINES Paris and was funded by ArcelorMittal Global R\&D, OCAS N.V.

\bibliographystyle{unsrt}

\end{document}